# NO$_2$ adsorption on GaN surface and its interaction with the yellow-luminescence-associated surface state

Yury Turkulets[1], Nitzan Shauloff[2], Or Haim Chaulker[1], Raz Jelinek[2], Ilan Shalish[1,*]

[1]*School of Electrical Engineering, Ben-Gurion University, Beer Sheva 8410501, Israel.*
[2]*Department of Chemistry, Ben-Gurion University, Beer Sheva 8410501, Israel*
*\*Corresponding author e-mail address: shalish@bgu.ac.il*

Trapping of charge at surface states is a longstanding problem in GaN that hinders a full realization of its potential as a semiconductor for microelectronics. At least part of this charge originates in molecules adsorbed on the GaN surface. Multiple studies have addressed the adsorption of different substances, but the role of adsorbents in the charge-trapping mechanism remains unclear. Here, we show that the GaN surface selectively adsorbs nitrogen dioxide (NO$_2$) existing in the air in trace amounts. NO$_2$ appears to charge the yellow-luminescence-related surface state. Mild heat treatment in vacuum removes this surface charge, only to be re-absorbed on re-exposure to air. Selective exposure of vacuum-annealed GaN to NO$_2$ reproduces a similar surface charge distribution, as does the exposure to air. Residual gas analysis of the gases desorbed during heat treatment in vacuum shows a large concentration of nitric oxide (NO) released from the surface. These observations suggest that NO$_2$ is selectively adsorbed from the air, deleteriously affecting the electrical properties of air-exposed GaN. The trapping of free electrons as part of the NO$_2$ chemisorption process changes the surface charge density, resulting in a change in the surface band bending. Uncontrollable by nature, NO$_2$ adsorption may significantly affect any GaN-based electronic device. However, as shown here, a rather mild heat treatment in vacuum restores the surface state occupancy of GaN to its intrinsic state. If attempted before passivation, this heat treatment may provide a possible solution to longstanding stability problems associated with surface charge trapping in GaN-based devices.

## I. INTRODUCTION

GaN is a semiconductor material that has revolutionized power electronics over the past two decades. Owing to their superior characteristics, AlGaN/GaN high electron mobility transistors (HEMT) practically displaced classic Si MOS-FETs in electric vehicle and RADAR applications. However, GaN-based transistors have several disadvantages, which limit their operational characteristics and hold them back from reaching their full potential. One of them, of specifically high significance, arises from a charge trapped in the GaN surface and interface states.[1] Undesired migration of this charge during transistor operation is one of the main contributors to the current collapse and RDS-ON issues in GaN HEMTs.[2,3] It has been suggested that this surface charge originates, at least in part, in external parasitic molecules adsorbed on GaN surfaces.[1,4] These adsorbed molecules seem to hold the key to a deeper understanding of GaN surface states and to the possible elimination of their effect on the GaN HEMT operation.

The effect of the surface states on transistor stability is neither new nor limited to GaN. Already at the beginning of the Si MOS-FET era, deep states at Si surfaces and oxide interfaces were realized to have a major impact on transistor reliability and performance.[5-7] Large charge concentrations trapped in these states and its dynamics caused instabilities in the transistor operation.[8] For Si, this problem was eventually solved by the improvement of SiO$_2$ growth methods, and by heat treatment in forming gas.[9-12] Both these processes passivate the Si dangling bonds, hence deactivating the associated trap states. Unfortunately, what works for Si does not necessarily work for GaN. GaN technology still suffers today from the same problems experienced at the commencement of Si technology, although the chemistry of surface adsorption on GaN is different from that of Si. Multiple studies have pointed out the outstanding ability of GaN surfaces to adsorb various air constituents from the environment.[13-20] Moreover, multiple designs of gas sensors based on GaN were proposed.[21-35] However, the physics behind GaN surface adsorption is still not clearly understood.

Reliable characterization techniques are required to recognize the chemistry and physics underlying the formation and charging of surface states in GaN. Photoluminescence (PL) spectroscopy is a powerful tool for characterizing deep levels, perhaps the most commonly used one. However, it is incapable of directly distinguishing occupied from unoccupied states. Surface photovoltage spectroscopy (SPS), in contrast, is definitely capable of this task.[36] Since SPS measures the photo-induced change in the surface band bending, its sensing is practically limited to variations in the surface charge density.[37] When combined, PL and SPS afford incomparable advantages for tracing the behavior of surface states and their charge occupancy. A combination of PL and SPS have been used previously to characterize the surface states in GaN.[38]

Foussekis et al. used SPS to study the effect of ambient conditions on the surface band bending of GaN. They have found that UV illumination in vacuum desorbs certain molecules from the surface, while the same illumination in air promotes their re-adsorption.[39] A later study by the same group has shown evidence of surface desorption in vacuum at elevated temperatures.[37] Another report by Chakrapani suggested that the adsorption of oxygen and water molecules on the GaN surface introduces a deep acceptor that affects the surface depletion region.[40]

We have recently shown evidence that the commonly observed GaN yellow luminescence (YL) is emitted from a surface state and that the charge occupying the related surface state is of external origin.[41] A comparison of surface charge distributions measured in air, vacuum, and after heat treatment in vacuum was used to show the dynamics of the charge occupying the related surface states. The charge density obtained from an SPS spectrum of the YL-related state in air reveals that the distribution of this surface state was only partially occupied by charge. When the same measurement was repeated following a mild heat treatment in vacuum, this charge was gone. A following exposure of the same GaN sample to air restored the original charge distribution. PL spectra obtained under similar conditions did not show any significant difference before and after the heat treatment, suggesting that only the charge occupancy of the state changed, but not the state itself. This observation suggests that the surface charge trapped in the YL-related state originates from external molecules adsorbed on the surface. Heat treatment in vacuum desorbs these molecules along with their charge, whereas re-exposure to air enables re-adsorption of the molecules and re-occupation of the surface state. However, the identity of the molecules has so far remained unknown. Here, we provide an evidence that exposes the chemical identity of these molecules as no other than nitrogen dioxide ($NO_2$). $NO_2$ is a common pollutant that exists in the air at ppm levels, and as we show here, is responsible for charging the YL-related surface state. We believe this new evidence is a crucial step in the search for passivation techniques for GaN surfaces.

## II. Experimental Details

A metal-organic chemical vapor deposited (MOCVD) 2.2 μm thick GaN epi-layer on c-plane sapphire was used in this study. The layer was Ga-face and was doped with both Si and Mg. The Mg doping compensated the n-type Si doping only partially, resulting in an overall n-type conductivity. SPS spectra were acquired using a Besocke Delta Phi Gmbh Kelvin Probe. Light from a 300 Watt Xe short-arc lamp was monochromitized by a Newport MS257 spectrometer and filtered using long-pass order-sorting filters. A computer-controlled slit was implemented to provide constant photon flux over the entire spectrum. The sample was illuminated for 5 minutes at each photon energy step before measuring the contact potential difference (CPD). The sample was left to relax in the dark for at least 48 hours prior to spectral acquisition to allow it to establish equilibrium. All SPS measurements were carried out inside a grounded stainless-steel chamber. Depending on the experiment, the chamber was evacuated to a high vacuum (base pressure less than $1 \times 10^{-5}$ mbar) or evacuated and then filled with specific gases to atmospheric pressure. The samples were mounted on a thermally controlled stage that maintained a constant temperature of 305 K during the SPS measurements. Heat treatment in vacuum was carried out by heating the sample stage to 450 K for 24 hours and letting it cool down to 305 K before SPS acquisition.

Residual gas analysis was carried out using a Pfeiffer QMS-220 residual gas analyzer (RGA) in an ultra-high vacuum chamber pumped using a Varian V70 turbomolecular pump and a SAES Getters Nextorr D100-5 getter/ion pump. Residual gas spectra were acquired after reaching a vacuum level of less than $1 \times 10^{-8}$ mbar. The RGA filament was turned on for at least 2 hours before the experiment. Oclaro BMU25A_915_01_R03 915 nm infrared laser operated at 4-Watt optical power was focused on the GaN surface to heat it during the RGA experiment. The light-beam diameter was expanded to cover 80 % of the sample surface. We preferred laser heating over the standard hot stage because of its ability to target the sample surface, thereby minimizing outgassing from the surrounding parts of the system. The sample was placed on a stage made of a Si wafer, which was thoroughly cleaned by solvents using the RCA1 and RCA2 cleaning processes.[42] Following the cleaning, the stage was baked in UHV at 300 °C for 3 h until the RGA signal originating from surface desorption of the Si diminished. Only negligible surface desorption was observed during the laser heat treatment of the stage alone, prior to the experiment.

## III. Results and discussion

Figure 1a shows how a mild heat treatment of GaN in vacuum desorbs the charge associated with the YL-related state. It confirms the occurrence of the same phenomenon reported previously also in the samples used here.[41] Exposure of the sample to air restores the charge density to its original condition. This indicates that one or more constituents of air are responsible for charging the traps.

To identify the specific constituent, we exposed a vacuum-annealed GaN sample to various gases. Primary air constituents, such as $N_2$, $O_2$, Ar, $H_2$, $CO_2$, and their combinations did not restore the YL-related charge. Water was the only major constituent of air that affected the surface charge distribution. Exposure to water vapor carried into the chamber using $N_2$ gas resulted in the surface charge distribution shown in Fig. 1b. However, the resulting charge distribution peaked at 2.42 eV instead of the 2.08 eV observed in the air-exposed sample. In the following experiments, we exposed the sample to various rare substances that may exist in air or process environment at ppm concentrations. These included exposure to $CH_4$, CO, $NO_2$, $NH_3$, methanol, ethanol, and isopropanol. Recovery of the surface charge was observed only on exposures to CO or $NO_2$. The charge distribution on the CO-exposed sample peaked at ~2.5 eV and its shape was somewhat similar to the distribution of the water-exposed sample (Fig. 1c). However, only exposure of the same sample to $NO_2$ recovered the original surface charge distribution of the YL-related state (Fig. 1d). The difference observed between the $NO_2$ and the

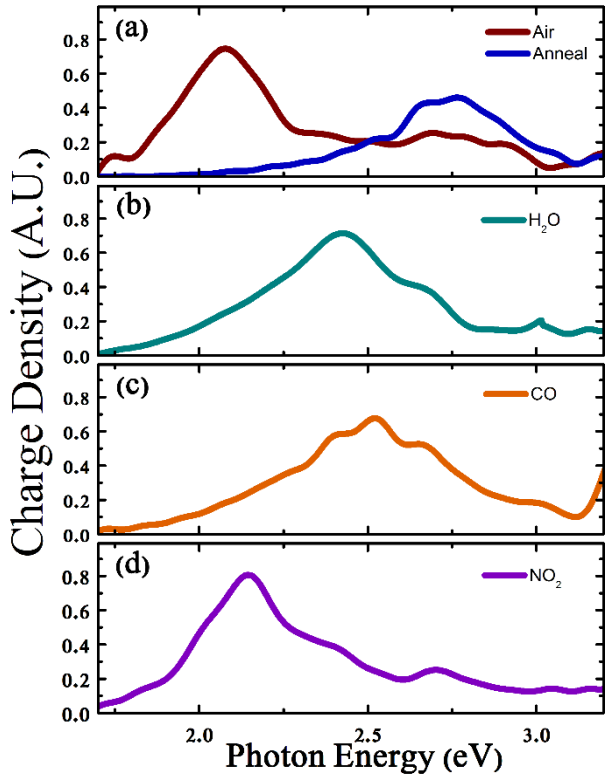

Fig. 1 Spectral distribution of the surface charge density of GaN (a) in air, and after the heat treatment in vacuum followed by exposure to $N_2$-diluted (b) water vapors, (c) CO, and (d) $NO_2$. Heat treatment in vacuum desorbs the surface charge, while exposure to air recovers it completely. Exposures of the sample to water vapor or CO restore the surface charge peaking at spectral locations different than the original YL-related peak. $NO_2$-exposure recovers the YL-related charge to its original peak position with a certain minor broadening of the peak to the blue parts of the spectrum.

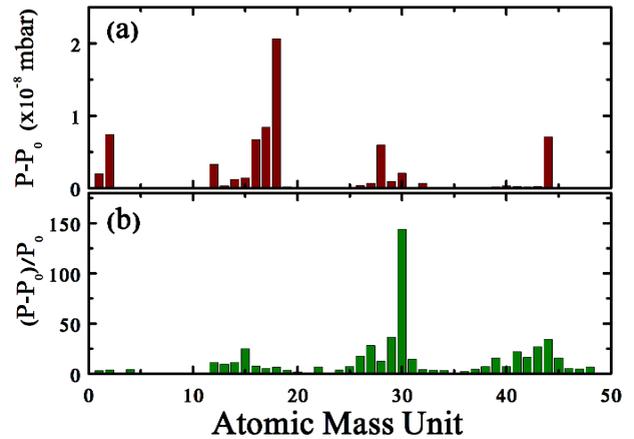

Fig. 2 (a) An RGA spectrum of substances vaporized from the GaN surface during heat treatment in vacuum. Except for the large $H_2$ peak, which is characteristic of turbomolecular pumped systems, the following gases (with their related AMUs) are observed: $H_2O$ vapors (18, 17), $CO_2$ (44, 28, 16, 12), CO (28, 16, 12), and NO (30). (b) The change factor of the partial pressure during the heat treatment relative to the partial pressure before the treatment is shown separately for each of the gasses. The NO-related peak stands out showing a 144-fold increase.

air-exposed results is a minor blue shift of 70 meV in the spectrum of the $NO_2$-exposed sample. This minor shift reflects the difference in the surface state occupancy between the air and the NO2 exposed samples and it means a slight difference in the surface band bending.[41]

To further validate the presence of NO2 adsorbates on a sample showing a YL-related charge, we carried out a residual gas analysis of the substance desorbed during the heat treatment in vacuum. The surface charge distribution of the air-exposed GaN was first acquired using SPS to ensure the presence of the YL-related charge. The sample was then annealed in an ultra-high vacuum (UHV), and the substances thus vaporized from its surface were analyzed using an RGA. The partial pressures of most of the present elements, as well as the total pressure in the chamber, peaked 50 sec after the laser was turned on. The RGA spectrum of the residual gas spectrum obtained at this point in time was used for the desorption analysis. To determine the composition of the gas vaporized from the GaN surface, the background RGA spectrum of the system (before the heat treatment) was subtracted from that obtained during the heat treatment (Fig. 2a). Water vapor was observed to be the main constituent at the atomic mass unit (AMU) 18 and 17. $CO_2$ was observed at AMU 44 and partially affected the peaks at 28, 16, and 12. However, the large peak magnitude at AMU 16 suggests a contribution of another component additional to $CO_2$, possibly atomic oxygen, O. Another prominent peak detected at AMU 28 was attributed to CO rather than $N_2$, owing to its contribution to the peak at AMU 12. The large peaks observed at AMUs 2 and 1 are explained by the major desorption/out-diffusion of H2 from GaN and by the reduced low-mass pumping speed of the turbomolecular pump. Finally, the peak observed at AMU 30 could, in principle, be attributed to either $NO_2$ or NO or both. However, the absence of an additional peak at AMU 46 overrules the presence of $NO_2$.[43]

To calculate the factor by which the partial pressures increased during the heat treatment, we divided the spectrum by the background spectrum measured before the heat treatment (Fig. 2b). The result shows an outstandingly large, 144-fold NO peak (at AMU 30) compared with the background spectrum. This change is exceptionally higher than the relative changes in the peaks of all the other gases. This observed NO emission is so uncharacteristic of the behavior of vacuum systems during a heat treatment that it provides bold evidence of massive outgassing of NO from the GaN surface.

Another notable observation is the minor increase in H2-related peaks. This suggests that the high $H_2$ concentration observed in the RGA spectrum is due to the poor pumping speed of hydrogen rather than an outgassing of the sample and may safely be regarded as an artifact. After subtracting the $H_2$-related peaks from the RGA spectrum, the composition of the desorbed gas was determined to be: H2O – 62.7 %, CO2 – 15.2 %, CO – 10.2 %, NO – 4.2 %, $O_2$ – 2 %. The remaining 5.7 % were mainly hydrocarbons. Hydrocarbons are common on air-exposed surfaces.[43]

Desorption of $H_2O$ during the heat treatment in vacuum could, in fact, originate at any part of the vacuum system. However, since only the sample surface was targeted by the focused laser beam, we believe that most of the desorbed water indeed originated on the sample surface. CO and $CO_2$ may also be generated by the RGA hot filament. However, this generation is not likely to show a significant increase during the heat treatment of the sample since the filament was turned on 2 hours before the experiment, and the background spectrum was subtracted from the spectrum obtained during the heat treatment. The NO molecule may also be produced by a reaction of N2 and $O_2$ on the hot RGA filament. However, this may only be possible in the presence of both $N_2$ and $O_2$ at high enough concentrations. Both these gases can desorb from the sample surface or otherwise result from an air leak. However, the partial pressures of $N_2$ and $O_2$ observed in our RGA spectra were too low to explain the high NO peak observed at AMU 30, let alone the 144-fold increase of this emission upon the heat treatment. An air leak is unlikely as well, in view of the relatively low Ar peak observed at AMU 40.

$H_2O$, CO, and $NO_2$ can all be absorbed on semiconductor surfaces, affecting their electrical properties through chemical or physical interactions. Surface adsorption of $H_2O$ is not surprising, as most semiconductor surfaces are known to be hydrophilic to a certain extent.[20] The selective adsorption of $NO_2$ may be attributed to its high free radical reactivity, possessing an unpaired electron that readily forms a bond and induces surface modifications.[44] In contrast, CO typically exhibits low reactivity on semiconductor surfaces, mostly by weak interactions. However, it has been theoretically shown that in the case of GaN, CO is readily adsorbed on the surface and behaves as an electron acceptor.[45]

Although we observe that all three species, $NO_2$, CO, and $H_2O$, are readily adsorbed on the GaN surface, only the $NO_2$ charge distribution spectrum actually correlates with the spectrum of the air-exposed sample. Moreover, the RGA results confirm excessive desorption of NO from the air-exposed sample. These results strongly suggest that the GaN surface selectively adsorbs $NO_2$ from the air. We will, therefore, limit the following discussion to the $NO_2$ adsorption mechanism.

Several studies have proposed a model for $NO_2$ adsorption on the surface of a GaN-based sensor.[35,46-48] All of these studies were unaware of any association of the NO2 with the YL-related surface state. A similar model has been proposed to explain the $NO_2$ sensing mechanism on surfaces of other semiconductors.[49-53] According to this model, being an electron acceptor, $NO_2$ captures an electron from the conduction band upon adsorption by the following reaction:

$$NO_2 (g) + e^- \rightarrow NO_2^- (ads) \qquad (1)$$

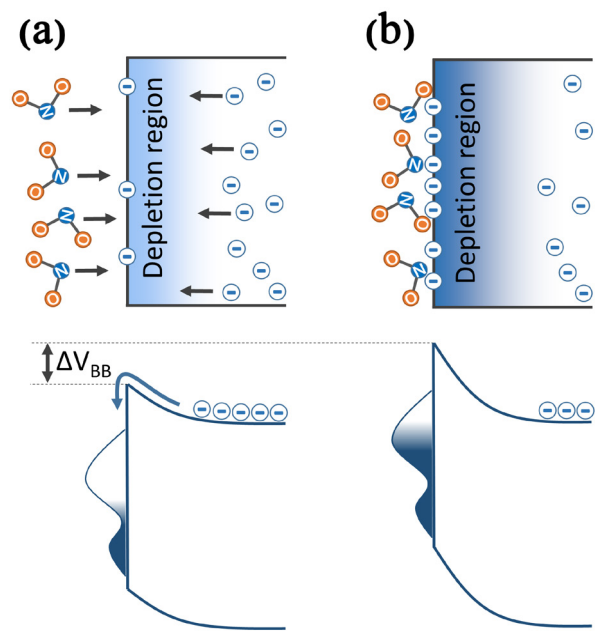

Fig. 3 Schematic illustration of NO₂ adsorption on GaN along with the corresponding band diagram change. (a) Before the adsorption the surface state is only partially occupied by electrons. NO₂ molecules that approach the GaN surface attract free electrons from the bulk into the surface state. (b) Free electrons, captured by the surface state, bind NO₂ molecules to the surface of GaN. The electron population in the surface state increases, leading to an increase in the surface band bending, detected as photovoltage. The desorption process occurs in a reversed order. Trapped electrons are excited by the thermal energy thus depopulating the surface state, releasing NO₂ molecules from the GaN surface. Consequently, the surface band bending decreases.

The captured electron charges the surface, increasing the surface depletion region and the surface band bending (surface voltage). Our method provides a previously-unknown detail of this process: this added surface charge appears to populate the YL-related surface state (Fig. 1). Figure 3 illustrates the dynamics of the adsorption process. When the GaN surface is exposed to $NO_2$, free electrons from the GaN bulk are attracted by the $NO_2$ molecules as they are adsorbed on the surface (Fig. 3a). The YL-related surface state provides a hosting site for the attracted electron that binds a $NO_2$ molecule to the GaN surface. Trapping of free electrons at the surface increases the surface state population, increasing the surface band bending (Fig. 3b). The inverse process, desorption, occurs in the opposite order. Thermal energy provided during the heat treatment excites electrons back into the conduction band, unbinding the $NO_2$ molecule, which becomes free to leave the GaN surface. Upon $NO_2$ desorption, the electron population in the surface state decreases, leading to a decrease in the surface band bending (measured as the surface photovoltage). This model suggests that the YL-related defect acts as a hosting site for an extrinsic $NO_2$ molecule, allowing its chemisorption on the GaN surface by capturing free electrons from the bulk.

From the electrostatic point of view, an adsorption/desorption mechanism of $NO_2$ on the GaN surface is not incidental. A free radical, $NO_2$ tends to satisfy its unpaired electron that originates from the nitrogen atom upon adsorption, forming a covalent bond with the surface.[54] To interact electrically, an adsorbent should act either as a donor or as an acceptor. An acceptor will withdraw an electron, while a donor will donate an electron to the surface. Either process works to minimize the electron energy. Thus, when an acceptor is adsorbed, it captures a conduction electron from the hosting semiconductor surface. When the adsorbent is a donor, it releases an electron to the valence band in a process commonly denoted as a hole capture. Figure 1 shows that in our case, the YL-related state is being populated by electrons during the adsorption process, suggesting that $NO_2$ acts as an acceptor capturing GaN conduction electrons upon adsorption. A surface defect localizes the captured electron at the surface, in immediate proximity to the impinging $NO_2$ molecule, making possible the formation of a covalent bond.[55]

Without $NO_2$, part of the free electrons from the n-type GaN bulk are trapped by surface states. These trapped electrons induce an electric field that creates a surface potential barrier, opposing further transport of electrons to the surface, and the system reaches a certain thermodynamic equilibrium. In this thermodynamic equilibrium, a balance is reached between the trapping of conduction electrons and thermal emission of trapped electrons over the potential barrier. However, this equilibrium is not static. Drift-diffusion processes, along with a thermionic emission over the barrier cause continuous movement of electrons back and forth between the bulk and the surface, although the average surface charge remains constant.[56] In our case, this surface charge density is observed as occupied surface states at the bottom part of the spectrum (blue curve in Fig. 1a).

When $NO_2$ comes into contact with the surface, it disrupts the existing equilibrium. An electron traveling from the bulk to the surface falls into an available surface state. Having a $NO_2$ molecule in its immediate proximity, the surface state strongly attracts this electron, creating a covalent bond with the $NO_2$. As more $NO_2$ molecules adsorb on the surface, the density of trapped electron charge increases, increasing the surface band bending (and the surface built-in field). This process continues until the system reaches a new thermodynamic equilibrium between the trapping process on the one hand and the repulsive electric field on the other hand, and the adsorption of additional $NO_2$ ceases. This new equilibrium has a higher density of electrons trapped at the surface, observed in our results as the building up of charge density within the YL-related energy range (red curve in Fig. 1a).[55]

Heat treatment increases the thermionic emission of electrons over the surface barrier back into the bulk. The emission of the bonding electron releases the $NO_2$ molecule, which is then pumped out of the vacuum system. The lower the concentration of adsorbed $NO_2$ at the surface, the lower the probability of an electron traveling from bulk to the surface to be re-trapped. Thus, the charge occupancy of the YL-related state gradually diminishes, and the original thermodynamic equilibrium is reestablished.

A possible role for surface defects in the $NO_2$ sensing mechanism has already been proposed for the case of a ZnO-based $NO_2$ sensor. Chen et al. used diffuse reflectance infrared Fourier transform spectroscopy (DRIFT), x-ray photoelectron spectroscopy (XPS), and PL to show that the oxygen vacancy, an intrinsic defect, captures an electron from the conduction band when ZnO is exposed to air.[57,58] In this process, one or two electrons are moved from the conduction band into the oxygen vacancies and from there into an adsorbed $O_2$ molecule. These captured electrons are later transferred to an adsorbed $NO_2$ to form a $NO_2^-$ ion. Upon temperature increase, adsorbed $NO2^-$ ions thermally decompose by releasing previously captured electrons. This proposed mechanism for $NO_2$ adsorption on ZnO incorporates intermediate electron transfer through an adsorbed $O_2$ molecule – an extra step that is not required in our model. Still, this model relies on an intrinsic surface defect to act as a hosting site for the free-electron-capture-induced surface adsorption of $NO_2$, as we propose for the GaN case.

So far, we have discussed $NO_2$ adsorption and desorption. The residual gas analysis suggests that the desorbed species is NO rather than $NO_2$. $NO_2$ appears to lose one of its oxygens upon desorption, and the question is, what is the mechanism underlying this decomposition upon heat treatment in vacuum?

Careful analysis of the RGA spectrum (Fig. 2a) shows that in addition to NO, there is a large signal at AMU 16, which may be attributed to atomic oxygen. Such a significant AMU 16 signal is not characteristic of the typical RGA spectrum. Typically, when molecular oxygen is detected, it appears in the RGA spectrum as a prominent peak at AMU 32 accompanied by a 9-fold weaker peak at AMU 16.[43] Detection of monoatomic oxygen alone is, therefore, not typical of oxygen desorption and is likely a product of decomposition of $NO_2$ to NO and O during the heat treatment. Thus, $NO_2$, adsorbed on the GaN surface as a complete molecule, decomposes by the thermal energy applied during the desorption process.

At atmospheric pressure, $NO_2$ is known to decompose thermally at temperatures as low as 150 ºC.[59] Rosser et Wise have shown that $NO_2$ decomposes into NO and $O_2$ at elevated temperatures by the following reaction[60]:

$$2NO_2 (g) \rightarrow 2NO (g) + O_2 (g) \qquad (2)$$

Since their experiment was carried out at an atmospheric pressure, the thermal decomposition of $NO_2$ exhibited a bimolecular gas-phase reaction. However, in ultra-high vacuum, the probability of two oxygen atoms meeting each other and forming molecular oxygen is extremely low due to their high mean free path of ca. $2 \times 10^6$ cm.[61] Therefore, in our experiment, $NO_2$ decomposed to NO and atomic O during the thermally-induced desorption process by the following reaction:

$$NO_2 (ads) \rightarrow NO (g) + O (g) \qquad (3)$$

Various electrical parameters of GaN-based devices have been reported to be affected by exposure of device surfaces to various gases.[62] These effects often lead to applications of GaN as a chemical sensor. Strikingly, it has been shown that GaN HEMTs are highly sensitive to various free radicals, and in particular $NO_2$.[21] Based on studies of GaN transistor $NO_2$ sensing, it has been suggested that the gate metal is involved in the sensing process as a catalyst dissociating $NO_2$ into NO and an oxygen ion.[22-26] Other studies involved the functionalization of AlGaN/GaN structure using various materials.[27-35] Recessed AlGaN/GaN open-gate structures were found to be highly sensitive in $NO_2$ detection.[30,31] Vitushinsky et al. showed that exposure of an open-gate structure to $NO_2$ decreased the 2DEG charge density. They hypothesized that being an electron acceptor, $NO_2$ captures electrons from non-ionized donors at the AlGaN surface, thereby increasing the positive charge at the surface. This positive surface charge decreases the 2DEG charge density.[30] Offermans et al. reported high selectivity of their sensor to nitrogen oxides compared with various other gases. Their sensor significantly responded only to NO and $NO_2$.[31] The clear benefits of an increased sensor surface area in gas sensing applications have also inspired the use of GaN nanowires and nanorods to detect $NO_2$.[32-34] Although these numerous reports implemented different sensor designs, they had one property in common – all of them were made of GaN. Hence, they lend strong support to our claim that $NO_2$ sensitivity is an intrinsic property of the GaN surface.

The high selectivity of the GaN surface to $NO_2$, found beneficial in sensing applications, has an opposite, undesirable effect on GaN-based power devices. When the device surface is exposed to the environment during its fabrication, it may adsorb $NO_2$ molecules undesirably. Any subsequent dielectric deposition step only buries these adsorbed molecules under it, making it impossible ever to passivate or neutralize their electrostatic effect. The charge trapped by those molecules easily migrates between the surface and the device's active area when the latter is biased, introducing instability in the device operation.[63-65] Moreover, variation in the $NO_2$ concentration in air may introduce an additional inconsistency in the resulting surface charge density. As shown here, a mild heat treatment in vacuum attempted prior to the passivation step may efficiently prevent these adverse effects and will ensure that the surface states of GaN are properly exposed to enable passivation by the deposited dielectric.[66]

## IV. Conclusion

This study rigorously establishes what appears to have been known for some time to GaN gas-sensor designers that GaN surfaces strongly react to, and are strongly affected by $NO_2$, an air pollutant of minute levels. The previously unknown physics underlying the mechanism of this interaction and its unequivocal association with the YL-related surface state are established here using a set of experiments that shed new light on an important mechanism of surface state charging in GaN. $NO_2$, selectively adsorbed from air, attracts free electrons to the GaN surface, where they become trapped by an intrinsic surface state. A heat treatment in vacuum is shown to reverse this process by depopulating the surface state and desorbing $NO_2$ molecules as NO and monatomic oxygen, O. Being an uncontrolled process, $NO_2$ adsorption may result in an unpredictable density of the surface charge, introducing uncertainty to future characteristics of GaN device. To prevent this uncontrollable surface charge, a heat treatment in vacuum may be attempted prior to the deposition of dielectric passivation.

**Acknowledgment**

Financial support from the Office of Naval Research Global through a NICOP Research Grant (No. N62909-18-1-2152) is gratefully acknowledged.